%Paper: gr-qc/9307021
%From: jasjeet@iucaa.ernet.in (Jasjeet Bagla)
%Date: Thu, 15 Jul 93 10:08:34 GMT

\magnification=\magstep1
\input paper
\voffset=-.5truein
\vsize=9truein
\baselineskip=14pt
\pageno=1
\pretolerance=10000
\def\n{\noindent}
\def\s{\smallskip}

\def\m{\medskip}
\def\c{\centerline}

\line{\hfil IUCAA-16/June'93}

\vskip 3cm

\c{\mid On Singularity Free Cosmological Models}

\vskip 1cm

\c{\bf N. Dadhich$^*$, R. Tikekar$^{**}$ and L.K. Patel$^{***}$}
\vskip 0.5cm

\c{\bf Inter-University Centre for Astronomy and Astrophysics}
%\vskip 0.5cm

\c{  Post Bag 4, Ganeshkhind, Pune - 411 007 , India.}
\vskip 0.5 cm

\c{\mid ABSTRACT}
\s

\n {\bf In 1990 Senovilla$^1$ obtained an interestisng cosmological solution of
Einstein's
equations that was free of  the big-bang singularity. It represented an
inhomogeneous and anisotropic cylindrical model filled with disordered
radiation,
${\bf \rho = 3p}$. The model was valid for ${\bf t \rightarrow - \infty }$ to
${\bf t \rightarrow
\infty}$ having all physical and geometrical invariants finite and regular
for the whole of spacetime. This was the first instance of a singularity free
cosmological model satisfying all the energy and causality conditions and
remaining
true to general relativity (GR). Subsequently a family of singularity free
models has been identified${\bf ^2}$. In this letter we wish to point out that
a simple and
natural inhomogenisation and anisotropisation, appropriate for cylindrical
symmetry,
of the Friedman-Robertson-Walker (FRW) model with negative curvature
leads to the same singularity free family. It consists of  the complete
set of singularity free
general solutions of Einstein's equations  for perfect fluid when cylindrically
symmetric metric
potentials are assumed to be separable functions of radial and time
coordinates.}
\vfill

\n $*$E-mail address : naresh@iucaa.ernet.in
\s
\settabs  4 \columns

\+$^{**}$Permanent address   &:Department of Mathematics, Sardar Patel \cr
\+&University, Vallabh Vidyanagar - 358120 (India) \cr
\s
\+$^{***}$ Permanent address  &:Department of Mathematics, Gujarat University,
\cr
\+&Ahmedabad - 380 009, (India).\cr

\vfill\eject

\n The standard Friedmann-Robertson-Walker cosmological model has been
quite successful in describing the present state of the Universe. It
prescribes a homogeneous and isotropic distribution for its matter content. It
is
though realised that homogeneous and isotropic character of spacetime cannot be
sustained at all scales, particularly for very early times. Furthermore, not
to have to assume very special initial conditions as well as for formation
of large scale structures in the Universe it is imperative to consider
inhomogeneity and anisotropy.

\s

\n The first step in this direction came in the form of the study of
anisotropic Bianchi models. Then inhomogeneity was also brought in and some
inhomogeneous models were considered$^{3-7}$. One of the main characteristics
of the Einsteinian  cosmology is the prediction of a big-bang singularity in
the finite
past. All these models (Bianchi as well as inhomogeneous) suffer from
singularity at $t = 0$. This experience was strongly aided by the general
result
that under physically reasonable conditions of positivity of energy, causality
and
regularity etc., the initial singularity is inescapable in cosmology so long as
we
adhere to Einstein's equations (singularity theorems$^8$). This gave
rise to the folklore that the big-bang singularity is the essential
property of the Einsteinian cosmology and it can only be avoided by
 invoking quantum effects and/or modifying Einstein's theory.

\s

\n On this background it was really refreshing when Senovilla$^1$ obtained
a new class of exact solutions of Einstein's equations without the
big -bang singularity. It represented a cylindrically symmetric Universe filled
with perfect fluid $(\rho = 3p)$. It was smooth and regular
everywhere, satisfied the energy and causality conditions and all the
physical as well as geometrical invariants remained finite and regular
for whole of spacetime.This marked the advent of singularity free cosmology.
It is important to recognise that the occurence of singularity is not
the general feature of Einstein's equations and for its avoidance it is not
always necessary to resort to quantum effects and other fields. The classical
Einstein's theory does admit cosmological models without sigularity
with physically acceptable behaviour for its matter content. It should
be noted that prior attempts to construct singularity free models had either
to ascribe physically unaceeptable behaviour for matter leading to violation
of energy and causality conditions or to invoke quantum effects or
 modification of GR$^{9, 10}$. Senovilla's$^1$ was the first singularity
free solution true to GR, conforming to energy and causality conditions.
Not only physical parameters are finite and regular, the solution has
been shown to be geodesically complete.$^{11}$ Physically it means that a test
particle will never encounter a singular state for arbitrarily large values of
its affine parameter. That is the particle trajectory will never terminate
anywhere.

\s

\n One may wonder how do these solutions escape singularity theorems$^8$? It is
because they do not satisfy one of the assumptions of the theorems, namely
existence of compact trapped surfaces$^8$. This assumption has always been a
suspect and does not appear as obvious and natural as the energy and
causality conditions. Violation of this means that nowhere in
spacetime gravity becomes strong enough to focus  fluid congruences in a
small compact region so that all particles including photons get trapped. The
occurence of such a situation appears natural for gravitational collapse
but by no means so for cosmology. For instance, even the open FRW model, that
has big-bang singularity, never encounters a trapped surface. Hence existence
of
trapped surfaces cannot be a natural property for cosmological models. All
previous attempts to construct singularity free models have tampered the energy
or causality conditions or GR. The remarkable feature of these models is that
they adhere to all physically acceptable conditions and avoid the application
of
singulairty theorems through non-existence of trapped surfaces.

\s

\n Ruiz and Senovilla$^2$ have separated out a fairly large class of
singulairty free models through a comprehensive study of a general
cylindrically symmetric metric with separable functions of $r$ and $t$. In this
note we wish to establish a link between the FRW model and the singularity free
family by deducing the latter from the former. It works like this: transform
the  FRW
metric with negative curvature into cylindrical coordinates and then introduce
inhomogeneity and anisotropy by pasting the  functions , that occur in FRW,
with
different powers in the metric coefficients. It is a simple and natural
inhomogenisation and anisotropisation process that leads to the singularity
free family.

\s

\n We begin with the FRW metric for the open Universe,
$$ ds^2 = dt^2 - T^2(t) \left( {dr^2 \over 1+ r^2} +
r^2 d\theta^2 + r^2 sin^2 \theta d\phi^2 \right)\eqno(1) $$

\n and transform it into cylindrical coordinates

$$ds^2 = dt^2 - T^2(t) \left( { d\bar r^{2} \over 1 + \bar r^{2}} +
(1 + \bar r^2 ) dz^2 + \bar r^2 d \phi^2 \right) \eqno(2)$$

\n by the transformation

$$r = \left( sinh^2z + \bar r^2 cosh^2z \right)^{1/2},~~ tan \theta = { \bar r
\over
sinhz \sqrt{1 + \bar r^2} }.
\eqno(3) $$
\n Further writing $m \bar r = sinh (m \hat r)$ and then dropping caps
to write

$$ ds^2 = dt^2 - T^2(t) \left( dr^2 + cosh^2 (mr) dz^2 +
m^{-2} sinh^2 (mr) d \phi^2 \right). \eqno(4) $$

\n Let us now inhomogenise and anisotropise the FRW metric by writing

$$ ds^2 = T^{2 \alpha} cosh^{2a} (mr) (dt^2 - dr^2) - T^{2 \beta}
cosh^{2b}(mr) dz^2 - m^{-2} sinh^2(mr) T^{2\gamma}cosh^{2c}(mr) d\phi^2
\eqno(5) $$

\n where we have used the coordinate freedom to write $g_{tt} = |g_{rr} |.$
We could have as well used the from (2). Taking the natural velocity field
${\bf u} = T^{\alpha} cosh^a(mr) dt,$ the isotropy of fluid uniquely determines

$$ T = cosh(kt), ~~ \alpha = \gamma .\eqno(6) $$

\n With this the metric (5) is the family of singularity free models identified
by Ruiz and Senovilla$^2$.

\s

\n Notice that $m^{-2} sinh^2 (m r) $ is simply to ensure $2 \pi$
periodicity for the angle $\phi$ and elementary flatness near the axis
and hence it does not participate in the inhomogenisation and anisotropisation
process. Ruiz and Senovilla$^2$ have taken $\beta + \gamma = 1 $ and
different undetermined functions of $r$ in place of $cosh (m r)$ and have
found that all functions are expressible as powers of the same function $cosh(m
r) $.
For time dependence $T = cosh(k t )$ is the general solution. Even if we take
$\beta + \gamma \not= 1$ and different $T(t)$ functions, it turns out that they
can all
be given as the powers of the single function, as given by (6). Thus the metric
(5)
with (6) forms the complete set of singularity free solutions of cylindrically
symmetric metric with separable functions of $r$ and $t$.

\s

\n For singularity free models, both Weyl and Ricci curvatures  should be
regular and their regularity for the metric (5) demands $\alpha = \gamma$.  The
isotropy of pressure constrains the parameters and it can be shown that the
only two following cases give rise to singularity free models :

\item{(i)}$ b =c, \alpha = \gamma,  \alpha + \beta = 1, ~ a = -b/(1+2b),
k = (1 + 2b)m ,$
\s

\item{(ii)} $b+c =1, \alpha = \gamma, \alpha + \beta = 1, a = -b (1 -b),
k = 2m.$

\n In the former case there does not occur an equation of state $\rho = \mu p$
in general, however for $b = - {1 \over 3}$ we obtain the Senovilla$^1$
radiation
model with $\rho = 3p$. In the latter case it is always $\rho = p$
giving the stiff matter model$^{12}$. The matter free limit $(\rho = 0) $ of
the stiff matter model yields two distinct singularity free vacuum solutions.
It may be noted that all these are the general solutions in the given setting.

\s

\n The kinematic parameters, expansion, shear and acceleration are given by

$ \theta = (\alpha + 1 ) k sinh (kt) cosh^{-\alpha - 1} (kt)
cosh^{-a}(mr)$

\s

$\sigma^2 = 2 (2 \alpha -1 ) k^2 sinh^2 (kt) cosh^{-2(\alpha + 1)}(kt)
cosh^{-2a} (mr)$
\s

$\dot u_r = - am  sinh(mr) cosh^{-a- 1}(mr)cosh^{-\alpha}(kt). $

\n It is clear that the above kinematic parameters are regular and finite all
through the spacetime. We have verified that so are the physical parameters
$\rho, p$ and the Weyl curvatures.  The general behaviour of the model is the
same as that of the Senovilla's radiation model. As $ t \rightarrow \pm \infty$
density and curvatures
tend to zero though the metric does not go over to the Minkowski from. The
Universe begins with low density at $t \rightarrow - \infty$, contracts to
a dense state at $t = 0 $(where density can be made as large as one pleases by
specifying the parameter $k$) and then starts expanding to reach the initial
state $\left(t \rightarrow - \infty \right)$ as $t \rightarrow \infty$. At $t =
0$ both expansion and shear change
their sense.

\s

\n It appears that presence of shear and acceleration seems to play an
important role in  avoidance of singularity. It is conceivable that they do not
let fluid
congruences to focus into  small enough a region to form trapped surfaces
leading
to singularity.  When fluid congruence has non-zero acceleration and there
occurs spatial {\it pressure gradient} which will counteract gravitational
attraction to give rise to a bounce to the model. This is how contraction
changes into expansion at $t = 0$ without letting the Universe to pass through
a singular state. The shearing of the congruence has a defocussing effect.
Their presence alone however is not sufficient to avoid singularity as there
exist singular models with non-vanishing shear and acceleration.
For instance replace $cosh(kt) $ in the above by $sinh(kt)$ to get a class of
models
with the big-bang singularity with shear and acceleration present. Thus it may
perhaps be the necessary condition but not sufficient. This added with the
regularity of Weyl and Ricci curvatures , and energy conditions may lead to
sufficiency. We have also verified that the metric(5) is geodesically
complete$^{13}$ for $\alpha \geq 0,  \alpha + \beta \geq
0,  \alpha \geq \beta, a \geq 0, a \geq b, a + b \geq 0, $ and $ b \leq 0 $.

\s

\n The metric (5) with (6) can as well be cast in the form

$$ \eqalign{ds^2 = &(1 + k^2t^2 )^{\alpha - 1 } ( 1 + m^2r^2)^a dt^2 - (1 +
k^2t^2)^{\alpha}
(1 + m^2r^2)^{a-1} dr^2 - (1 + k^2t^2)^{\beta}\cr &(1 + m^2r^2)^b dz^2 - r^2
(1+k^2t^2)^
{\alpha} ( 1 + m^2r^2)^c d \phi^2 \cr}(7) $$

\n which reduces to the FRW from (2) for $\alpha=\beta=1, a=c=0$ and $b = 1 $
where
$T(t) = (1 + k^2t^2)$.
\n It is interesting that if one just uses hyperbolic or $(1 + k^2t^2)$
functions,
which are clearly the obvious choice for singularity free spacetime, one ends
up with the family of singularity free models. However the cylindrical symmetry
seems to
play an important role. The above prescription does not obviously work in
spherical symmetry. A tentative prescription for a singularity free spacetime
may be as follows:
Use non-singular elementary functions  having no zeros as metric potentials,
and ensure regularity of Weyl and Ricci curvatures with non-zero shear and
acceleration.This is just the broad setting, the fluid consistency conditions
and energy conditions are then to be satisfied. This prescription does yield
acceptable fluid models for
cylindrical symmetry.

\s

\n One may ask the question, how robust is the singularity free framework in
relation
to accommodating other force fields ? It turns out that viscosity cannot be
included without sacrificing positivity of viscosity coefficients for all
time$^{14}$ while
the radial heat flow can easily be included.$^{15}$ Both the cases above can be
generalised to have radial heat flow. Note that for $\rho = \mu p$, $\mu$ can
have the only two discrete values $(\mu = 1, 3)$. If we introduce massless
scalar
field alongwith perfect fluid in the case (i), the resulting fluid can have an
equation of state , $4 > \mu > 3,$ opening out a narrow window for $\mu$.

\s

\n Finally the most pertinent question for the singularity free models
is: how to evolve them into the (present day) FRW models? The question is
inherently very difficult because the former are cylindrically symmetric
wherein
a direction is singled out while the latter are spherical having no
identifiable direction.  It may be noted  that the ratio of shear to
expansion, that measures the anisotropy, is a constant for these models.We have
here
 established a linkage between the two which may
help in reconciling them together. The affirmative answer to this question will
have very important bearing on our overall cosmological perception of the
Universe and
in particular for the early Universe cosmology. The other question is, do there
exist general solutions without any symmetry or is cylindrical symmetry singled
out for
singularity free solutions ?

\s

\n Details will be given in the forthcoming paper$^{16}$ .
\s

\n Acknowledgements: LKP and RST thank IUCAA for hospitality.

\vfill\eject

\n{\bf References}

\m

\item{1.}  Senovilla, J.M.M. Phys. Rev. Lett. {\bf 64}, 2219, (1990).

\s
\item{2.}  Ruiz,  E.\&  Senovilla, J.M. Phys. Rev. D 45, 1995 (1992).

\s

\item{3.}  Wainwright, J. \&  Goode, S.W. Phys. Rev. {\bf D 22}, 1906 (1980).

\s
\item{4.} Feinstein, A. \& Senovilla,J.M.M. Class Quantum Grav. {\bf 6}, 189
(1989).
\s
\item{5.} Davidson,W.  J.Math. Phys. {\bf 32}, 1560, (1990).
\s

\item{6.} Patel, L.K.  \& Dadhich, N. Ap. J., {\bf 401}, 433 (1992).
\s
\item{7.} Patel,  L.K. \& Dadhich, N. J. Math. Phys. (To appear)

\s
\item{8.}Hawking,  S.W. \&  Ellis,G.F.R. The Large Scale Structure of the
Universe (Cambridge University Press, 1973).

\s

\item{9.}  Murphy, J.M. Phys.Rev. D8, 4231, (1973).

\s
\item{10.} Berkenstein,  D. \& Meisels, A. Astrophys. J. {\bf 237}, 342 (1980).

\s
\item{11.} Chinea, F.J.   Fernandez-Jombrina ,L.\&  Senovilla,J.M.M. Phys. Rev.
{\bf D45}, 481 (1992).
\s
\item{12.} Patel,  L.K.  \& Dadhich, N. Preprint:IUCAA-1/93
\s
\item{13.} Dadhich, N. \& Patel,L.K.  under preparation.
\s
\item{14.}  Patel, L.K.\& Dadhich, N. Preprint: IUCAA-21/92
\s
\item{15.} Patel , L.K.  \& Dadhich, N. Preprint: IUCAA-2/93 to appear in
Class. Quant.
Grav.
\s
\item{16.}  Dadhich, N.  Patel, L.K. \&  Tikekar, R. under preparation.

\bye